\documentclass[reprint,superscriptaddress, 
 amsmath, amssymb, aps, prb,  
 ]{revtex4-2}

\usepackage{graphicx}
\usepackage{dcolumn}
\usepackage{amsmath,amssymb,bm}
\usepackage[none]{hyphenat}
\usepackage{verbatim}
\usepackage{textcomp}
\usepackage{subcaption}
\usepackage{hyperref}
\hypersetup{colorlinks=true, urlcolor=blue, linkcolor=blue, citecolor=blue}
\begin{document}

\preprint{}

\title{Effect of Ti-doping on the dimer transition in Lithium Ruthenate}

\author{Sheetal~Jain}
\author{Zhengbang~Zhou}
\author{Ezekiel~Horsley}
\author{Christopher~J.~S.~Heath}
\affiliation{University of Toronto, Toronto, Ontario, M5S 1A7, Canada}
\author{Mohsen~Shakouri}
\author{Qunfeng~Xiao}
\author{Ning~Chen}
\author{Weifeng~Chen}
\author{Graham~King}
\affiliation{Canadian Light Source (CLS), Saskatoon, Saskatchewan, S7N 2V3, Canada}
\author{Young-June~Kim}
\email{youngjune.kim@utoronto.ca}
\affiliation{University of Toronto, Toronto, Ontario, M5S 1A7, Canada}

\date{\today}

\begin{abstract}
We carried out a comprehensive crystal structure characterization of Ti-doped lithium ruthenate (Li$_2$Ti$_x$Ru$_{1-x}$O$_3$), to investigate the effect of Ti-doping on the structural phase transition. Experimental tools sensitive to the average structure (X-ray diffraction), as well as those sensitive to local structure (Extended X-ray Absorption Fine Structure, EXAFS; pair distribution function, PDF) are used. We observed non-monotonic dependence of the structural transition temperature on the Ti-doping level. At low doping, the transition temperature slightly increases with doping, while at high doping, the temperature decreases significantly with doping. We note two important observations from our studies. First, Ti K-edge EXAFS data shows persistent Ti-Ru dimerization even with substantial Ti doping. Second, we were able to use the PDF data to estimate the dimer correlation length above the transition temperature, which would correspond to the size of the proposed local `dimer clusters' formed by Ru-Ru and Ti-Ru neighbours. The dimer correlation length is found to be around 10~\AA, which remains robust regardless of doping. Our study therefore suggests that Ti$^{4+}$ with its $d^0$ electronic configuration is a special type of dopant when replacing Ru.
\end{abstract}


\maketitle

\clearpage


\section{\label{sec:intro}Introduction}

Many honeycomb lattice materials are drawing attention in recent years due to their interesting magnetic properties. For example, transition metal halides, such as CrI$_3$, have been extensively investigated due to their two-dimensional ferromagnetic properties~\cite{Huang2017}. At the same time, Kitaev's honeycomb model and Jackeli and Khaliullin's prescription for bond-dependent interaction~\cite{JK2009} have inspired many investigations into honeycomb magnets with strong spin-orbit coupling (SOC), such as Na$_2$IrO$_3$~\cite{Singh2010}, $\alpha$-Li$_2$IrO$_3$~\cite{Singh2012,Freund2016}, H$_3$LiIr$_2$O$_6$~\cite{Kitagawa2018}, and $\alpha$-RuCl$_3$~\cite{Kim2022}. In this context, Li$_2$RuO$_3$ is a curious exception among honeycomb magnets. The $d^4$ electronic configuration of Ru$^{4+}$ in an octahedral crystal field would be expected to have $S=1$ local moment in the absence of SOC. Experimentally, however, Li$_2$RuO$_3$ is non-magnetic at room temperature, which may be explained by strong SOC and the resultant $J_{eff}=0$ state of the Ru$^{4+}$ ion. What is surprising is that a paramagnetic phase turns on sharply above $T_c\approx 550$~K.

\begin{figure*}[htbp]
\centering
\begin{subfigure}[b]{0.472\textwidth}
\includegraphics[width=\linewidth]{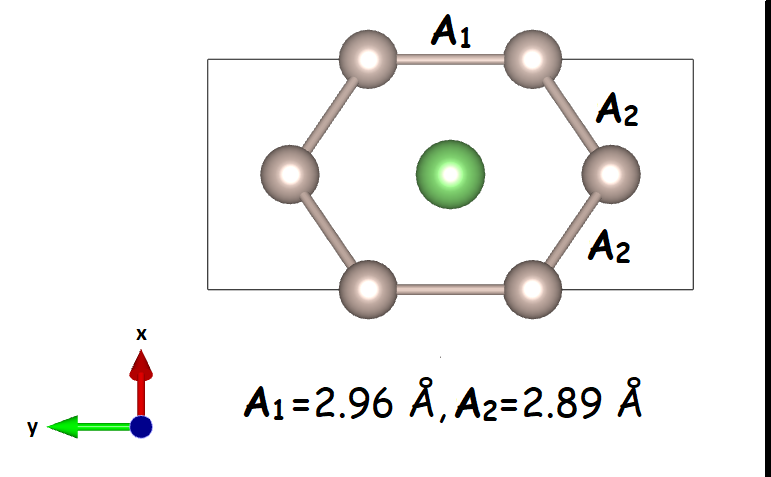}
\caption{\label{fig:s1}$C2/m$ structure}
\end{subfigure}
\begin{subfigure}[b]{0.52\textwidth}
\includegraphics[width=\linewidth]{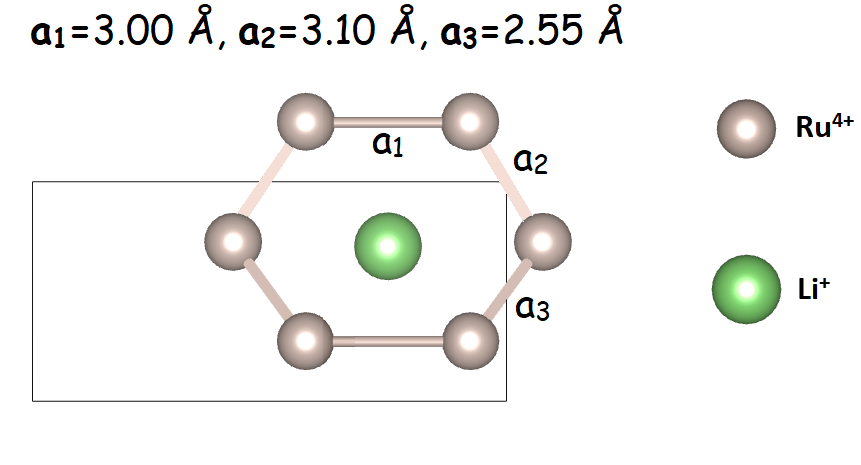}
\caption{\label{fig:s2}$P2_1/m$ structure}
\end{subfigure}
\caption{\label{fig:structure}Schematic representation of the in-plane Ru-Ru neighbours in crystal structure of Li$_2$RuO$_3$, for \subref{fig:s1}~$C2/m$ and \subref{fig:s2}~$P2_1/m$ structure. The monoclinic unit cell is shown with the solid black line in both cases. Crystal structure visualization was done using the VESTA software~\cite{Momma2011}.}
\end{figure*}

This observation was explained via a careful structural study by Miura and coworkers~\cite{Miura2007}, who showed that the room temperature structure is monoclinic $P2_1/m$, with a significantly distorted hexagon of Ru-Ru neighbours in-plane. This is visualized in Fig.~\ref{fig:s2}. Here, the shortest Ru-Ru neighbour distance, $a_3$, is referred to as a `dimer', and the longest distance, $a_2$, is called the `inter-dimer' distance. The remaining bond length $a_1$ is close to the undimerized neighbour distance. At high temperature (600~K), the diffraction pattern matched that of $C2/m$ structure with more even Ru-Ru bond length distributions as shown in Fig.~\ref{fig:s1}~\cite{Miura2007}.

It was proposed that the 4d $t_{2g}$ orbitals in Ru-Ru neighbours hybridized to form molecular orbitals, resulting in a dimer formed via direct Ru-Ru bonding. Various first-principles calculations support this picture~\cite{Miura2009,Jackeli2008,Johannes2008}. 
However, subsequent study of the local crystal structure of Li$_2$RuO$_3$~\cite{Kimber2014} showed that the dimers and the local distortion persisted well above the transition temperature. The interpretation of this surprising observation was that the in-plane Ru-Ru hexagon was distorted and `dimerized' at all temperatures, with the dimers simply being ordered below the transition, and disordered at higher temperatures. The disordered dimers {\em on average} can be described by `global' crystal structure with the $C2/m$ symmetry. The phase transition thus becomes of order-disorder type, and Kimber et al. termed the low temperature phase a valence-bond (dimer) solid, and the high temperature phase a valence-bond liquid~\cite{Kimber2014}. 

While this valence bond physics is intriguing, and has been extensively studied ~\cite{Wang2014,2016,Park2016,2016-2,Mehlawat2017,Arapova2017,Zheng2019,Yun2019,Han2020,P2015,Reeves2019,Ponosov2019,Ponosov2020,2022,Bansal2020,Bansal2022,Hoang2019}, the robust dimerization prevents one from observing exotic quantum magnetism predicted for a $4d^4$ system with strong SOC. For example, the singlet $J_{eff}=0$ ground state exchange-coupled to low-lying $J_{eff}=1$ triplet state could give rise to excitonic magnetism ~\cite{Khaliullin2013}. In a recent attempt to suppress the dimerization, Takayama et al. studied Ag intercalated Li$_2$RuO$_3$ under high pressure, which revealed complex magnetic phases when intermediate pressure was applied~\cite{Takayama2022}.
Another way to disturb the dimerization tendency is to frustrate dimer ordering by diluting Ru sites by replacing Ru with different metal ions. In fact, several studies have already tried using Mn, Ir, or Li to suppress the dimer ordering structural transition~\cite{Yun2021,Lyu2017,Mori2016,Lei2014}. However, the dopant ions used in earlier studies can be considered a ``strong" perturbation since they are either magnetic (Mn or Ir) or charged (Li). Structurally, these ions also have very different ionic radii -- Mn$^{4+}$ (67 pm), Ir$^{4+}$ (82 pm), and Li$^{+}$ (90 pm) -- from that of Ru$^{4+}$ (76 pm) ~\cite{Shannon1976}.  

The goal of this paper is to study the effect of replacing Ru with non-magnetic and isovalent Ti ions in Li$_2$RuO$_3$. In comparison to Mn/Ir/Li, Ti$^{4+}$ (74.5 pm) has ionic radius similar to that of Ru$^{4+}$. Since Ti$^{4+}$ is very similar to Ru$^{4+}$, one would naively expect the dilution to be a good way to investigate the effect of the frustrated dimer state without the local strain caused by dissimilar size of dopants. We note that previous studies of Ti-doping of Li$_2$RuO$_3$ were focused exclusively on electrochemical performance characterization, with very limited crystal structure characterization~\cite{Sathiya2015,Kalathil2015,Zhao2019,Moradi2020,Yao2021,Terasaki2015}. Our main focus here is the impact of replacing Ru with Ti on the structural phase transition, characterized by both X-ray diffraction (XRD) to study the `global' crystal structure, and EXAFS (Extended X-ray absorption fine structure) \& PDF (pair distribution function) to study the `local' crystal structure. We employed two types of local structure probes in this study, which provide complementary information. While EXAFS provides higher ``spatial" resolution to distinguish different bond lengths, the PDF method can be used to probe extended local correlation beyond the nearest-neighbors.

To our surprise, we found that a small amount of Ti dopants actually enhance the structural transition temperature, indicating that they do not disturb the locally dimerized state. Only when a substantial amount of Ti is introduced to the system, we observe suppression of dimerization. We discuss this surprising observation in the context of covalent bonding between Ti and Ru, which is not common in complex oxide structures. Our study reveals that Ti dilution provides a unique way to investigate the dimer liquid phase in the presence of quenched disorder.


\section{\label{sec:methods}Experimental methods}

\subsection{Sample Preparation}

Polycrystalline (powder) samples of undoped and Ti doped Li$_2$RuO$_3$ (Li$_2$Ti$_x$Ru$_{1-x}$O$_3$) for low doping levels of $x=\{0.05, 0.10, 0.12, 0.15, 0.20, 0.30, 0.40, 0.50\}$ were synthesized via solid state reaction~\footnote{We use the term doping instead of dilution to describe our samples, since dilution is often used in the context of diluting magnetic elements. As described above, Ru is non-magnetic in pure Li$_2$RuO$_3$ and magnetic properties are not studied here.}. The raw materials used were Li$_2$CO$_3$ (Alfa Aesar, 99.998\%), RuO$_2$ (Thermo Fisher Scientific, 99.9\%) and TiO$_2$ (Alfa Aesar, 99.9\%) powders. The raw materials were first separately dried at 623 K for 6 h to remove moisture. Then, stoichiometric quantities of each compound were mixed and ground thoroughly in an agate mortar and pestle. The mixtures were then pressed into 16mm-diameter pellets and heated at 1273 K for 48 h, with intermediate grindings and re-pelletization. A 10\%-mol excess of dried Li$_2$CO$_3$ was added during the heat treatment, to compensate for lithium deficiency due to the volatility of Li$_2$O at the synthesis temperature (1273 K)~\cite{James1988,P2015,Zhao2019,Yun2021,Lei2014,Bansal2022,Bansal2020}. A previous study on undoped lithium ruthenate found that the principal impurity in this synthesis is RuO$_2$, and its absence ensures high purity~\cite{2016}. Thus, in our doped sample, the final product was inferred to be crystallographically pure Li$_2$Ti$_x$Ru$_{1-x}$O$_3$ after confirming no trace of RuO$_2$. The sample oxidation state remains in Ti$^{4+}$ and Ru$^{4+}$ for all the samples studied here, as confirmed by our X-ray absorption spectroscopy (XAS) data shown in 
the Supplemental Material~\cite{[{See Supplemental Material at }]supp}.

\subsection{X-ray Diffraction}

All XRD measurements were done using a Rigaku Smartlab diffractometer. The powder samples were thoroughly ground before the measurement, and mounted in Bragg-Brentano geometry. Cu K$\alpha$ radiation was used, with Cu K$\beta$ blocked using a nickel filter. Heating and cooling from room temperature were done using Anton Paar DHS and DCS sample stages respectively, with the sample in vacuum ($<$10$^{-1}$ mbar).  

Rietveld refinement of XRD data of all samples was done using the GSAS-II software~\cite{Toby2013}. Rietveld refinement produced fits for the data and background, and yielded the lattice constants and unit cell volume of the monoclinic unit cell, as well as isotropic mean-square displacements ($U_{iso}$) for each atom. Note that $U_{iso}$ was computed instead of anisotropic thermal parameters to obtain values with reasonable error bars. 

\begin{figure*}[hbtp]
\centering
\includegraphics[width=0.99\textwidth]{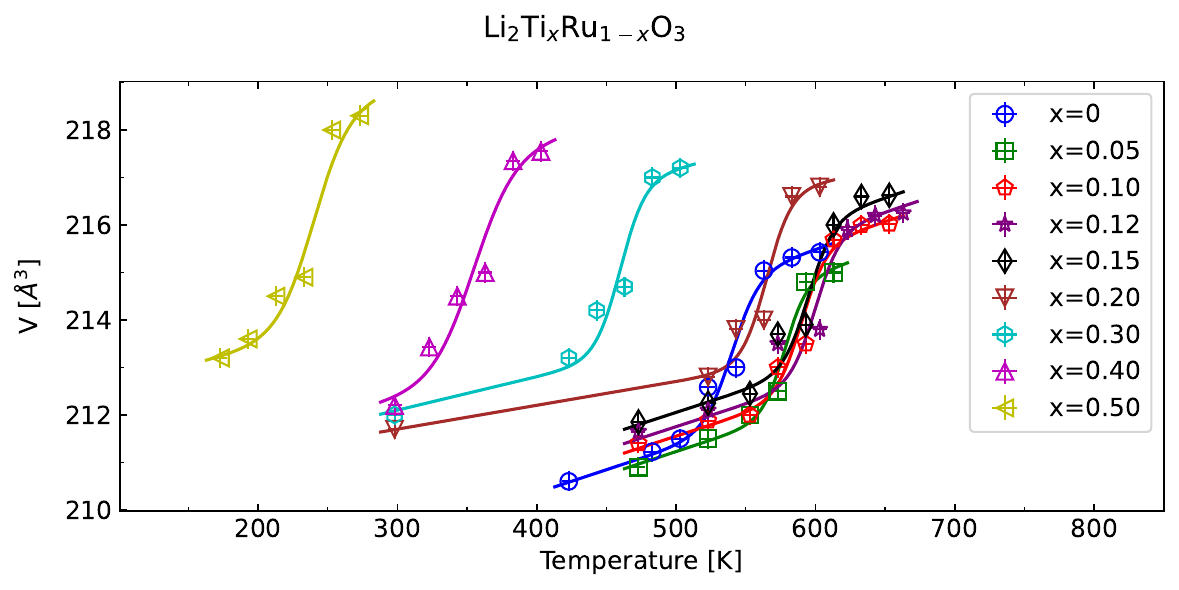}
\caption{\label{fig:XRD1}~Temperature dependence of unit cell volume of Li$_2$Ti$_x$Ru$_{1-x}$O$_3$, obtained by Rietveld refinement, for various doping fractions ($x$). The bold lines, used to infer phase transition temperature ($T_c$), are fits to Eq.~\eqref{eq:eq1}.}
\end{figure*}

\subsection{EXAFS}
All EXAFS measurements were carried out at the Canadian Light Source (CLS). Ti K-edge (4966 eV) data was collected at the Soft X-Ray Microcharacterization Beamline (SXRMB), while Ru K-edge (22117 eV) measurements were conducted on the Hard X-ray Micro-Analysis (HXMA) beamline. Titanium K-edge data was collected only at room temperature, for all doped samples. For Ti K-edge measurements, powder samples were mounted onto a double-sided, conductive carbon tape and loaded into a vacuum chamber with $\sim$10$^{-7}$ torr vacuum, to optimize flux for lower photon energy. These measurements were performed in 90 degree geometry, with the sample being 45 degree to the detector and 45 degree to the incident beam.  A 7-element Silicon Drift Detector (SDD) was used to record the fluorescence yield (FLY). The total electron yield (TEY) was also recorded by measuring the drain current of the sample. On the other hand, Ru K-edge data was collected in transmission geometry.
For Ru K-edge, each sample was diluted with pure boron nitride powder, finely ground, pressed into a uniform pellet, and then affixed to polyimide tape. These procedures were done in order to achieve suitable element concentration and uniformity. Ru K-edge data at room temperature was collected for all samples (undoped and doped). For temperature dependence of Ru K-edge, select low ($x=0.10$) and high doping ($x=0.30$) samples were chosen, along with the undoped sample. The energy of the incident X-ray beam from the wiggler was selected using a double-crystal Si(111) monochromator, with higher harmonic contributions suppressed by a combination of Rh-coated mirrors and a 50\% detuning of the wiggler. For both edges, standard Ti and Ru foils were utilized for energy calibration. 
For all measurements, at least three scans were taken at each edge for each sample in order to guarantee data reproducibility and good signal-to-noise ratio. 

EXAFS data processing and fitting were done on the ATHENA and ARTEMIS software of the Demeter suite, respectively~\cite{Ravel2005}. In data processing, EXAFS data obtained in energy-space was reduced using standard procedures, and the resulting $k$-space data was Fourier transformed (F.T.) to $r$-space, yielding peaks that correspond to different shells of neighboring atoms in the EXAFS equation:\\
\begin{equation}\label{eq:eq2}
\chi(k) = \sum\limits_j \frac{S_0^2 N_j e^{-2k^2\sigma^2_j}e^{-2r_j/\lambda(k)}f_j(k)}{kr_j^2}\sin(2kr_j + \delta_j(k))
\end{equation}~\\
where $j$ represents different coordination shells made up of $N_j$ identical atoms at approximately the same distance from the absorbing atom; $S_0^2$ is the amplitude reduction factor which corrects for multi$-$electron scattering and other experimental effects (generally between 0.7 and 1.2); $\sigma^2_j$ is the atom-pair mean-square displacement along the atom-pair distance $r_j$; $f_j(k)$ and $\delta_j(k)$ are the scattering factor and scattering phase shift, which are provided in the FEFF program~\cite{Rehr1995}; $\lambda(k)$ is the mean-free-path of the photoelectron; $k = \sqrt{2m(E-E_0)/\hbar^2}$, where $E_0$ is the theoretical Fermi level position. In data fitting, the $r$-space data is fit to a sum of complex phase and amplitude functions calculated using the FEFF program, to extract atom-pair distances ($r_j$) and thermal factors ($\sigma^2_j$) for each atom-pair. Another two fitting parameters, $S_0^2$ and $\Delta E_0$ (the difference in edge energy between the value defined for the data and the theoretical function), are also determined for each edge. In this analysis, both the real and imaginary parts of the $r$-space data were fit.

\subsection{PDF}

All PDF data was collected using the Brockhouse X-ray Diffraction and Scattering - High Energy Wiggler (BXDS-WHE) beamline at the CLS. Incident energy of 60.8 keV was selected using Si (422) single crystal monochromator (side bounce) in Laue mode. The beam size was 0.05 $\times$ 0.2 mm. Samples were mounted in powder form in 0.5 mm diameter quartz capillaries and heated using a coil heater. A Varex (XRD 4343CT) scintillation area detector mounted at 179 mm from the sample was used to measure the data. For background reference, an empty capillary was measured and the resulting pattern was then subtracted from the data. The background-subtracted raw data sets were then radially integrated using the GSAS-II software~\cite{Toby2013}. PDF model fitting was performed using PDFGui~\cite{Farrow2007}, with the model calculations convoluted using a sinc function, to better represent the experimental $Q$-range of $1<Q<22$ \textup{\AA}$^{-1}$.

\section{\label{sec:results}Experimental results}

\subsection{\label{sec:xrd}X-ray Diffraction}

In order to characterize the effect of Ti-doping on the phase transition temperature($T_c$), X-ray diffraction (XRD) measurements were carried out as a function of temperature. The temperature dependence of the unit cell volume for all samples is presented in Fig.~\ref{fig:XRD1}. Note that there is a significant jump in the unit cell volume across the transition, as is known to be the case for the undoped sample~\cite{Miura2007,Kimber2014}. This jump is also observed for all the doped samples. The position of this jump in unit cell volume is used to estimate the phase transition temperature ($T_c$) for all samples. See the Supplemental Material~\cite{[{See Supplemental Material at }]supp} for refined structural parameters and also for the discussion of complementary methods used to determine the transition temperature.

The temperature dependence of the unit cell volume is fit to the following equation: \\
\begin{equation}\label{eq:eq1}
V(T) = V_0(1+\beta T) + V_1\;S(\;(T-T_c)/b\;)
\end{equation}~\\
where $V$ is the unit cell volume at temperature $T$, $V_0(1+\beta T)$ is a linear function to account for thermal expansion, $\beta$ is the volumetric thermal expansion coefficient and $S(x) \equiv 1/(1+e^{-x})$ is the sigmoid function, to `smoothly' model the jump across the transition. $V_1$ is the height of this jump, while $b$ is a phenomenological parameter to estimate the temperature spread of the transition about $T_c$. Fits based on Eq.~\eqref{eq:eq1} are shown via solid lines in Fig.~\ref{fig:XRD1} for each doping.

\begin{figure}[htbp]
\centering
\includegraphics[width=0.49\textwidth]{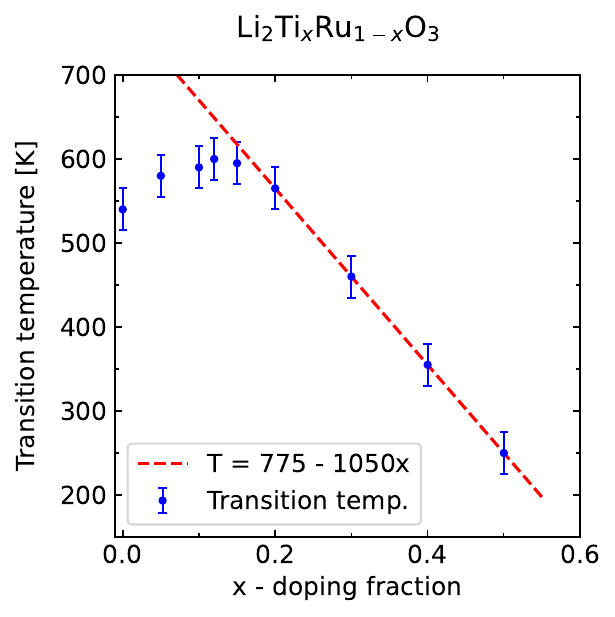}
\caption{\label{fig:XRD2}~Doping dependence of phase transition temperature ($T_c$) in Li$_2$Ti$_x$Ru$_{1-x}$O$_3$, obtained by fitting temperature dependence of unit cell volume. The dashed line is a linear fit, for high doping fractions ($x \geq 0.2$).}
\end{figure}

The transition temperature ($T_c$) for each doping, obtained from the fits described above, are plotted in Fig.~\ref{fig:XRD2}. Since the $b$ parameter describes roughly the half-width of the jump, it is taken to be the associated errorbar. $T_c$ shows non-monotonic behaviour as a function of doping, while $b$ remains more or less unchanged. The transition temperature for the undoped sample is $\sim$$550$~K, consistent with previous reports~\cite{Miura2007,Kimber2014}. The transition temperature shows a slight increase to $\sim$$600$~K for $x \leq 0.15$, after which the transition temperature decreases linearly for $x \geq 0.2$. The linear fit for decrease in transition temperature at high doping is shown via the dashed line. 

This is a surprising result, since one naively expects monotonic behaviour from the perspective of percolation. However, in the case of Ti-doping of Ru lattice site in Li$_2$RuO$_3$, the doping dependence of $T_c$ shows different, and almost contrasting, behaviour at low and high doping concentrations. Since this transition is due to dimer ordering (at least for x=0), we carried out local structure investigation for doped samples using two separate methods, EXAFS and PDF, as detailed in the next subsection. 

\begin{figure*}[htbp]
\centering
\begin{subfigure}[b]{0.49\textwidth}
\includegraphics[width=\linewidth]{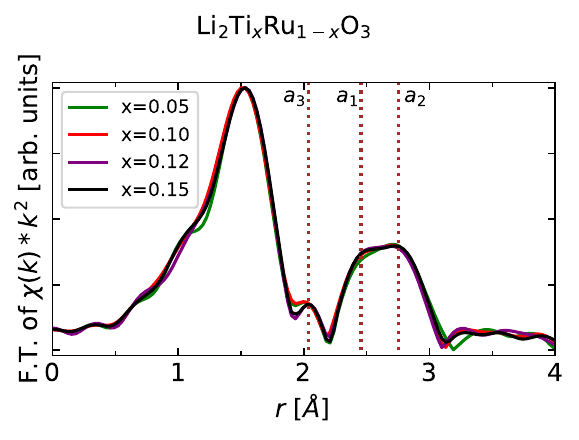}
\caption{\label{fig:Ti1}}
\end{subfigure}
\begin{subfigure}[b]{0.49\textwidth}
\includegraphics[width=\linewidth]{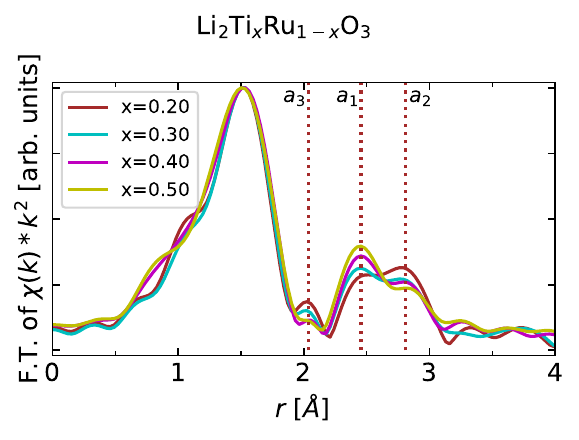}
\caption{\label{fig:Ti2}}
\end{subfigure}
\caption{\label{fig:Ti}Ti K-edge EXAFS data in $r$-space (normalized) for various doped samples of Li$_2$Ti$_x$Ru$_{1-x}$O$_3$ at room temperature, \subref{fig:Ti1}~$x=0.05$ to $x=0.15$ (low doping) and \subref{fig:Ti2}~$x=0.20$ to $x=0.50$ (high doping). The large peak around 1.5 \textup{\AA} corresponds to the Ti-O bond. The vertical brown dotted lines represent the first, second  and third (Ti-(Ru/Ti)) neighbour locations in $r$-space, corresponding to Ti occupying Ru-site in the crystal structure. The Fourier transform (F.T.) ranges are 3$-$12 \textup{\AA}$^{-1}$.}
\end{figure*}

Another interesting observation is the doping dependence of the unit cell volume. As seen in Fig.~\ref{fig:XRD1}, the average unit cell size increases with doping, which is consistent with results from a previous electrochemical study~\cite{Zhao2019}. This is most clearly visible for higher doping samples ($x>0.15$). Considering that Ti$^{4+}$ (74.5 pm) and Ru$^{4+}$ (76 pm) have similar ionic radii~\cite{Shannon1976}, this is somewhat surprising, and perhaps indicating the complex role Ti plays in replacing Ru in this material. 

\subsection{\label{sec:exafs}EXAFS}

\subsubsection{Ti K-edge}

Titanium K-edge EXAFS measurements were carried out at room temperature on all doped samples, since only the doped samples contain Ti. The EXAFS data in $r$-space is shown in Fig.~\ref{fig:Ti}. Note that the $r$-space peak positions are slightly different from the exact atom-pair (neighbour) distances, due to phase shift factor in the EXAFS equation, Eq.~\eqref{eq:eq2}. 

The first peak near 1.5 \textup{\AA} corresponds to the Ti-O bond, which suggests that Ti occupies the Ru-site, forming TiO$_6$ octahedra in the crystal structure. This Ru-site occupation is corroborated by the other three peaks, indicated by the vertical brown dotted lines. These correspond to the three neighbour distances in $r$-space, indicated by $a_3$, $a_1$ and $a_2$ respectively, following Fig.~\ref{fig:s2}. They correspond to the distance between Ti and the 3 neighbouring metal ions (Ti or Ru). For low doping concentrations, this will most likely be Ti-Ru bonds. A few salient features can be observed in this data. 

First, the peak corresponding to $a_3$ (short dimer distance) is present in all doped samples. This is unexpected, since the formation of dimers is attributed to the molecular-orbit formation between Ru ions~\cite{Miura2009,Jackeli2008}. This suggests that the Ti$^{4+}$ ion form a dimer with a neighboring Ru$^{4+}$ ion. This comes out from the experimental data, but will be discussed further in Sec.~\ref{sec:discuss}. 

\begin{figure}[htbp]
\centering
\includegraphics[width=0.5\textwidth]{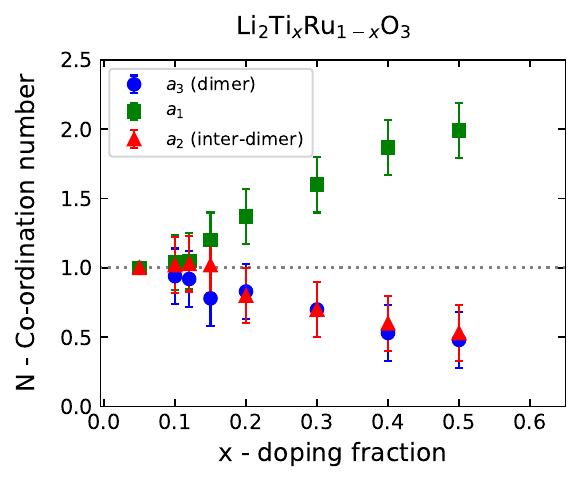}
\caption{\label{fig:Ti-N}Doping dependence of coordination number of the three (Ti-(Ru/Ti)) neighbours, obtained by fitting the Ti K-edge EXAFS spectra. Note that the changes are not significant at low doping ($x=0.05$ to $x=0.15$), and are very pronounced at higher dopings ($x \geq 0.20$). At high doping, Ti prefers to form $a_1$-type neighbours, reducing the number of $a_3$ (dimer) and $a_2$ (inter-dimer) type.}
\end{figure}

\begin{figure*}[htbp]
\centering
\begin{subfigure}[b]{0.49\textwidth}
\includegraphics[width=\linewidth]{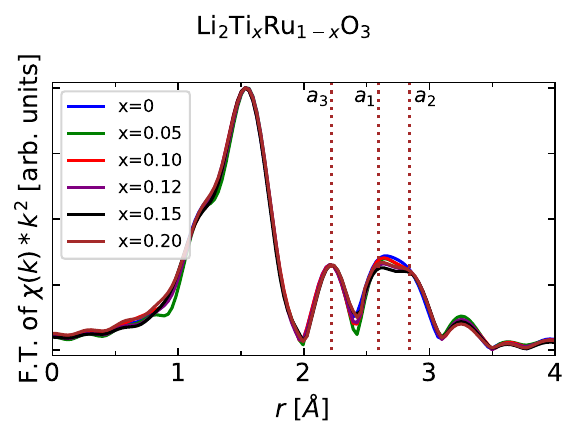}
\caption{\label{fig:Ru1}}
\end{subfigure}
\begin{subfigure}[b]{0.49\textwidth}
\includegraphics[width=\linewidth]{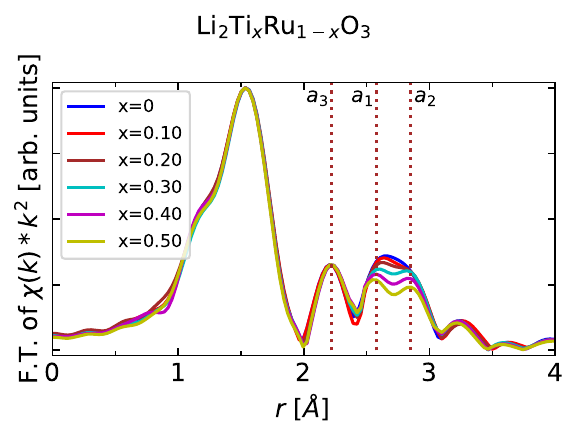}
\caption{\label{fig:Ru2}}
\end{subfigure}
\caption{\label{fig:Ru}Ru K-edge EXAFS data in $r$-space (normalized) for all samples of Li$_2$Ti$_x$Ru$_{1-x}$O$_3$ at room temperature, \subref{fig:Ru1}~$x=0$ to $x=0.15$ (low doping trend) and \subref{fig:Ru2}~$x=0$ to $x=0.50$ (high doping trend). The large peak around 1.5 \textup{\AA} corresponds to the Ru-O bond. The vertical brown dotted lines represent the first, second  and third (Ru-(Ru/Ti)) neighbour locations in $r$-space. The Fourier transform (F.T.) ranges are 3$-$14 \textup{\AA}$^{-1}$.}
\end{figure*}

Secondly, for $x < 0.2$, as shown in Fig.~\ref{fig:Ti1}, the data shows no significant change with doping, and in particular, the three (Ti-(Ru/Ti)) neighbour peaks have similar relative intensities (`heights'). However, for $x \geq 0.2$, these relative intensities show visible changes with doping, as seen in Fig.~\ref{fig:Ti2}. As doping concentration increases, the peak corresponding to $a_1$ increases in intensity, and the other two peaks, representing $a_3$ and $a_2$, decrease in relative strength. Since the relative intensity of EXAFS peaks is related to the co-ordination number, this suggests that the coordination number changes for the three (Ti-(Ru/Ti)) neighbours for $x \geq 0.2$. 

To obtain quantitative information, all EXAFS spectra were fit using the dimerized Li$_2$Ti$_x$Ru$_{1-x}$O$_3$ crystal structure, $P2_1/m$, with Ti dopant occupying the Ru-site in the crystal structure. Details of the EXAFS fitting are included in the Supplemental Material~\cite{[{See Supplemental Material at }]supp}. 
In the context of EXAFS, the coordination number refers to the number of distinct neighbours (usually of the same element type) present at the same distance in the 3D crystal structure, resulting in a multiplicative increase of the corresponding peak at the $r$-location in the EXAFS spectra. For example, the Ti-O bond peak near 1.5 \textup{\AA} has a coordination number of 6, since there are 6 oxygen neighbours at the same distance from Ti in the TiO$_6$ octahedra. The coordination number for this bond remains 6 for all dopings, since the Ti-O peak shows no splitting or appreciable changes in relative intensity.

In the case of the (Ti-(Ru/Ti)) neighbours, the fits yield coordination number of the three (Ti-(Ru/Ti)) bond lengths, and the doping dependence is plotted in Fig.~\ref{fig:Ti-N}. Titanium has all three neighbours in equal number at low doping ($x \leq 0.15$), with coordination number $N_1,N_2,N_3 \approx 1$. However, as doping increases, $N_1$ increases, reaching $N_1 \approx 2$ for $x=0.50$, while $N_2$ and $N_3$ concomitantly decrease. Note that the sum of the three coordination number is always 3, as expected (assuming no vacancies). This result suggests that, with increasing doping, the tendency for Ti to form a dimer decreases, presumably due to the fact that the probability for a Ti to find a Ti neighbor increases. In a crude approximation, for $x=0.50$, about half of the Ru-Ru dimers will be replaced by Ti-Ti pairs. If we assume that Ti-Ti prefers non-dimerized bond, roughly one half of $a_3$ (and corresponding amount of $a_2$) will take on a neutral bond length of $a_1$. This can explain $N_1 \sim 2$, and $N_2 \sim N_3 \sim 0.5$ for $x=0.50$. 

\subsubsection{Ru K-edge}

Ru K-edge EXAFS measurements were also performed in order to provide complementary information from the Ru point of view. The room-temperature data in $r$-space is shown in Fig.~\ref{fig:Ru}. 
Again, the first peak near 1.5 \textup{\AA} corresponds to the Ru-O bond of the RuO$_6$ octahedra in the crystal structure. Also, same notation for $a_3$, $a_1$ and $a_2$ is used. 
At low doping, the data shows no significant change with doping, shown in Fig.~\ref{fig:Ru1}, similar to the Ti case. The three (Ru-(Ru/Ti)) neighbour peaks have similar relative intensities. However, the high doping behaviour, shown in Fig.~\ref{fig:Ru2}, departs from that of Ti K-edge in Fig.~\ref{fig:Ti2}. The three (Ru-(Ru/Ti)) neighbour peaks actually retain the same coordination number of $N$$\approx$$1$ for all dopings. In addition, the peaks corresponding to $a_1$ and $a_2$, which are almost indistinguishable for lower doping samples, seem to become more resolved with increasing doping levels. This is apparent with the emergence of the dip between the two dotted lines near 2.7 \textup{\AA} in Fig.~\ref{fig:Ru2}. 
This suggests that the difference in neighbour distance of $a_1$ and $a_2$ increases with doping, at higher doping. This effect was also subtly seen in the Ti case at high doping in Fig.~\ref{fig:Ti2}. In contrast, the dimer distance $a_3$ does not change much, and remains the same within error bar for all samples, in both Ti and Ru EXAFS data. 
More details of quantitative EXAFS fits can be found in the Supplemental Material~\cite{[{See Supplemental Material at }]supp}. This result is consistent with the Ti EXAFS data shown in the previous subsection. For Ru$^{4+}$ ion, the $a_1$, $a_2$, $a_3$ bond local structure is fundamentally unchanged, regardless of whether the neighbor is Ru or Ti. This finding reinforces the idea that only Ti-Ti bond is responsible for modifying the low-temperature structure. 

\begin{figure*}[htbp]
\centering
\begin{subfigure}[b]{0.49\textwidth}
\includegraphics[width=\linewidth]{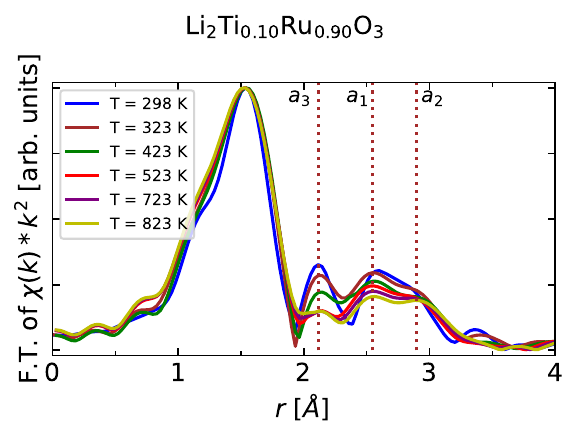}
\caption{\label{fig:Ru-T1}}
\end{subfigure}
\begin{subfigure}[b]{0.49\textwidth}
\includegraphics[width=\linewidth]{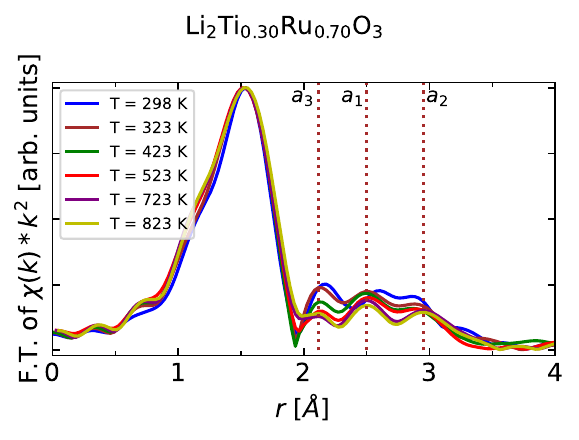}
\caption{\label{fig:Ru-T2}}
\end{subfigure}
\caption{\label{fig:Ru-T}Temperature dependent Ru K-edge EXAFS data in $r$-space (normalized) for \subref{fig:Ru-T1}~Li$_2$Ti$_{0.10}$Ru$_{0.90}$O$_3$ (low doping) and \subref{fig:Ru-T2}~Li$_2$Ti$_{0.30}$Ru$_{0.70}$O$_3$ (high doping). The dimers are seen to persist well above the transition temperature ($T_c$) for both samples. While some temperature dependence is observed in the peak positions (neighbour distances), the coordination number remains the same for all neighbours as a function of temperature ($N$$\approx$$1$). The Fourier transform (F.T.) ranges are 3$-$12 \textup{\AA}$^{-1}$.}
\end{figure*}

\begin{figure*}[htbp]
\centering
\begin{subfigure}[b]{0.49\textwidth}
\includegraphics[width=\linewidth]{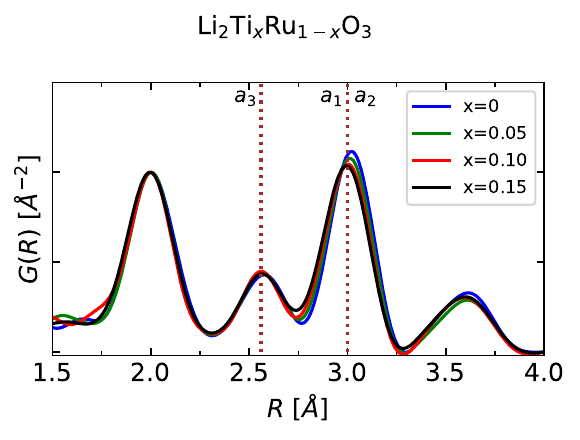}
\caption{\label{fig:PDF-RT-R1}}
\end{subfigure}
\begin{subfigure}[b]{0.49\textwidth}
\includegraphics[width=\linewidth]{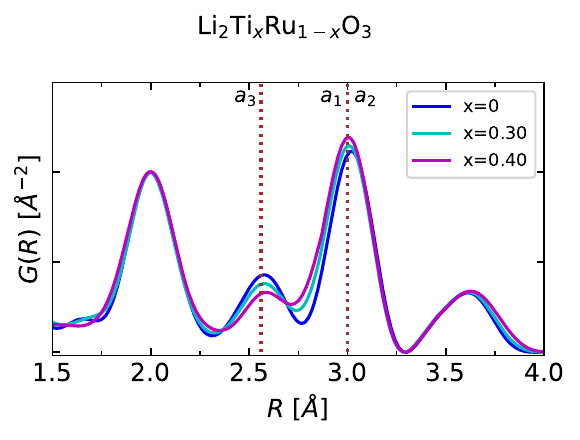}
\caption{\label{fig:PDF-RT-R2}}
\end{subfigure}
\caption{\label{fig:PDF}PDF data in $R$-space for various samples of Li$_2$Ti$_x$Ru$_{1-x}$O$_3$ at room temperature, \subref{fig:PDF-RT-R1}~$x=0$ to $x=0.15$ (low doping trend) and \subref{fig:PDF-RT-R2}~up to $x=0.50$ (high doping trend). The peak around 2 \textup{\AA} corresponds to the (Ru/Ti)-O bond. The vertical brown dotted lines represent the first and second/third (merged) Ru-site neighbour locations in $R$-space in the crystal structure. }
\end{figure*}

Temperature-dependent EXAFS measurements at Ru K-edge were conducted on the doped samples, and this data is presented in  Fig.~\ref{fig:Ru-T}. The results for $x=0.10$ are shown in the left panel, Fig.~\ref{fig:Ru-T1}, while the right panel has the EXAFS data for $x=0.30$, Fig.~\ref{fig:Ru-T2}. For both cases, peaks at the $a_3$ position are observed for even the highest temperatures, well above the structural phase transition temperature. This indicates that dimers persist locally above the phase transition temperature for the doped samples as well. Note that the apparent changes in the peak intensity are a result of changes in mean-square displacement along each neighbour, which leads to a broadening of the peaks. Fitting parameters can be found in the Supplemental Material~\cite{[{See Supplemental Material at }]supp}. We observe that the mean-square displacement of $a_3$ increases more than $a_2$ with increasing temperature, which in turn increases more than $a_1$.

\begin{figure*}[htbp]
\centering
\begin{subfigure}[b]{0.49\textwidth}
\includegraphics[width=\linewidth]{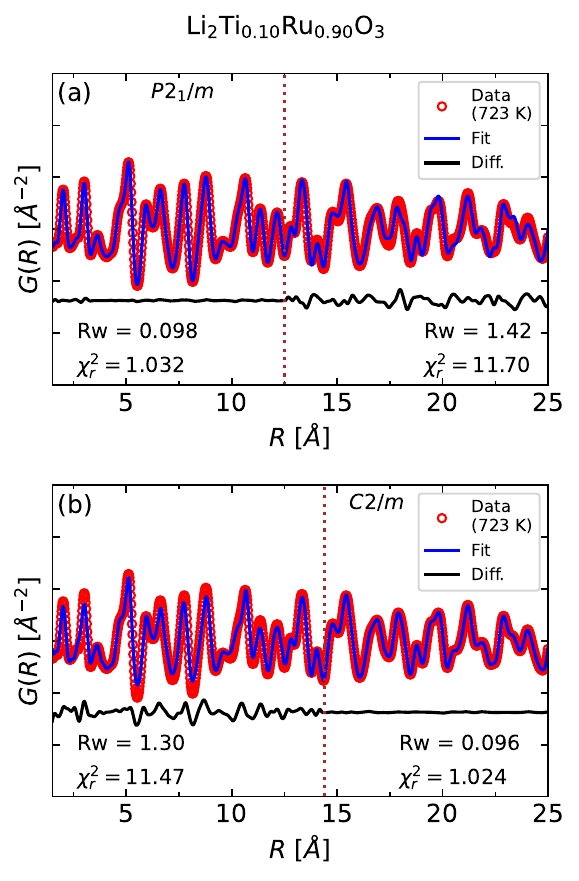}
\phantomsubcaption
\label{fig:PDF-fit-R1}
\phantomsubcaption
\label{fig:PDF-fit-R2}
\end{subfigure}
\begin{subfigure}[b]{0.49\textwidth}
\includegraphics[width=\linewidth]{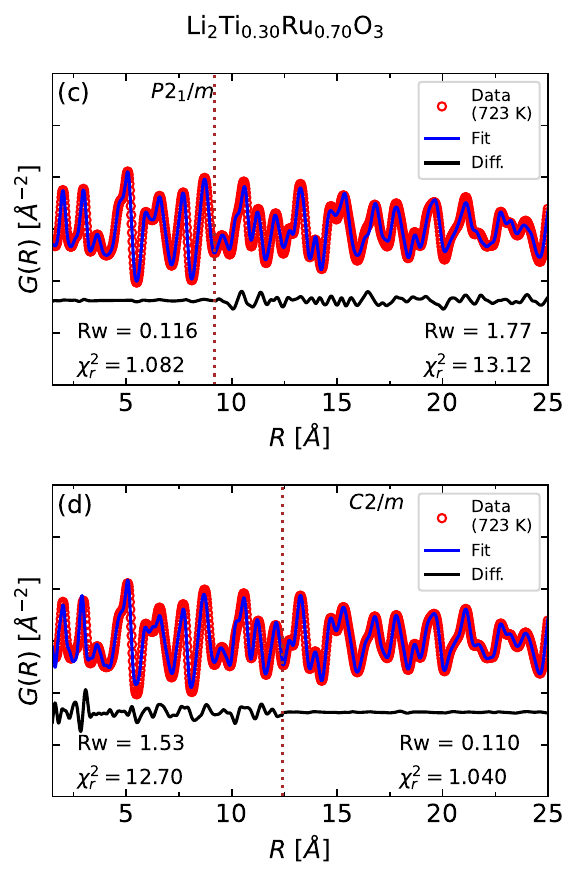}
\phantomsubcaption
\label{fig:PDF-fit-R3}
\phantomsubcaption
\label{fig:PDF-fit-R4}
\end{subfigure}
\caption{\label{fig:PDF-fit}Fitting of PDF data in $R$-space at 723~K for Li$_2$Ti$_{0.10}$Ru$_{0.90}$O$_3$ ($x=0.10$, low doping) using \subref{fig:PDF-fit-R1}~distorted $P2_1/m$ structure and \subref{fig:PDF-fit-R2}~undistorted $C2/m$ structure, as well as for Li$_2$Ti$_{0.30}$Ru$_{0.70}$O$_3$ ($x=0.30$, high doping), with \subref{fig:PDF-fit-R3}~$P2_1/m$ structure and \subref{fig:PDF-fit-R4}~$C2/m$ structure. The low $R$-range fits well for the dimerized structure ($P2_1/m$) and the high $R$-range to the undistorted structure ($C2/m$) for both samples. The vertical brown dotted lines represent the lower and upper bound of the dimer-correlation length in $R$-space in the crystal structure for each sample. }
\end{figure*}

On fitting the Ru K-edge EXAFS spectrum quantitatively, the coordination number obtained is the same for all neighbours at all temperatures ($N \approx1$), indicating that the local structure is preserved at all temperatures. This result is consistent with the earlier observation for the undoped sample that about 1/3 of all bonds are dimers, which go through the order-disorder transition at $T_c$~\cite{Kimber2014,Park2016}. 

\subsection{\label{sec:pdf}PDF}

Complementary information on local structure can be obtained from PDF measurements for undoped and select doped samples. While we lose the element-sensitivity of EXAFS, PDF allows us to gain quantitative information about slightly longer-range structural correlation as described below.
The room temperature results for PDF are shown in Fig.~\ref{fig:PDF}. The peak at 2 \textup{\AA} corresponds to the (Ru/Ti)-O bond, and its height (coordination number) is seen to be the same for all dopings. This is expected, since all samples have (Ru/Ti)-O$_6$ octahedra in the crystal structure, with the same coordination number of 6. The neighbours of the Ru-site, $a_3$, $a_1$ and $a_2$ are indicated by the vertical brown dotted lines. Note that, in the case of PDF, the peaks corresponding to $a_1$ and $a_2$ cannot be distinguished, due to the lower resolving power of PDF as compared to EXAFS. 

The doping dependence for $x=0$ to $x=0.15$ is shown in the left panel, Fig.~\ref{fig:PDF-RT-R1}. Note that the data shows no significant doping trend at low doping, indicating that the local environment of Ru-site ($a_3$, $a_1$ and $a_2$) remains the same. The high doping trend, in Fig.~\ref{fig:PDF-RT-R2}, shows a clear change with doping, where the peak associated with $a_3$ decreases and the merged peak of $a_1$ and $a_2$ increases with doping. This is expected, given that Ti prefers to have $a_1$-type neighbours, as seen in the Ti EXAFS results. 


The PDF data obtained at high temperatures above the phase transition also shows local dimerization, consistent with the EXAFS data shown in Fig.~\ref{fig:Ru-T}. Since the dimerization is a local distortion that is washed away when the average (`global') crystal structure is probed using XRD, the short-distance structure can be expected to closely resemble the low-temperature distorted structure. The PDF method allows us to examine the short ``correlation" range in which such a locally distorted structure remains a valid description. The PDF data would fit well with the distorted (i.e., dimerized) $P2_1/m$ structure in the short-range, but will become progressively worse at longer range, and vice versa for the undimerized $C2/m$ structure. We compare the fitting done using both the distorted structure ($P2_1/m$) and the undistorted structure ($C2/m$) for the PDF data obtained at temperatures above the transition ($T=723$K). In Fig.~\ref{fig:PDF-fit}, The PDF data is shown in hollow red circles, and the fits in solid blue lines. The black line at the bottom is the difference profile. The distorted $P2_1/m$ structure fits the data well for the low $R$-range, as seen in the top panels, Fig.~\ref{fig:PDF-fit-R1} and Fig.~\ref{fig:PDF-fit-R3}. The undimerized $C2/m$ structure, on the other hand, fits well for the high $R$-ranges, displayed in Fig.~\ref{fig:PDF-fit-R2} and Fig.~\ref{fig:PDF-fit-R4} for each sample respectively. In contrast, the low temperature data can be fitted using the $P2_1/m$ structure for the entire $R$-range as shown in the Supplemental Material~\cite{[{See Supplemental Material at }]supp}.

\begin{figure}[htbp]
\centering
\includegraphics[width=0.5\textwidth]{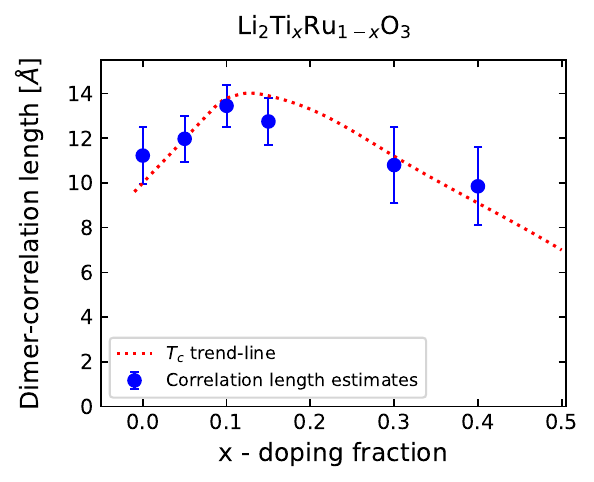}
\caption{\label{fig:PDF-corr}Doping dependence of dimer-correlation length estimates, obtained by fitting the PDF data at 723 K. Note that the bounds increase at low doping, till $x=0.15$ and decrease at higher dopings.}
\end{figure}

One can estimate how far in $R$ the local distortion remains correlated from the change in the goodness of the fit. For example, in Fig.~\ref{fig:PDF-fit-R1}, the residual value (Rw) is close to zero for $R < 12.5$~\AA, and reduced-$\chi^2$ close to 1, but both Rw and $\chi^2$ become much worse beyond this value. Similar boundaries can be found in the other panels. We define dimer correlation length as the average between the lower bound found in the $P2_1/m$ fitting and the upper bound found in the $C2/m$ fitting with the difference as error bar in our estimate. Note that, for low doping ($x=0.10$) in the left panels, the dimer-correlation length seems to be slightly longer than that of the high doping ($x=0.30$) case. These estimates are plotted in Fig.~\ref{fig:PDF-corr} as a function of doping. The dimer-correlation length remains almost unchanged (or slightly increases with doping) between $x=0$ to $x=0.10$ (low doping), and then decreases at higher dopings. The changes are small though, considering the large uncertainty in determining such a quantity. We note, however, that the doping trend shows some similarity with the transition temperature $T_c$, shown in Fig.~\ref{fig:XRD2}. To highlight this similarity, the scaled $T_c$ trend-line is superposed on the data as the red dotted line in Fig.~\ref{fig:PDF-corr}. The correlation length estimate $\approx 10$ \AA $ $ for the undoped sample matches results from a previous study~\cite{Kimber2014}. 

\section{\label{sec:discuss}Discussion}

Two significant observations in our study are the non-monotonic dependence of the structural transition temperature on the doping level and the persistent dimerization even with substantial Ti doping. We believe that these two observations are closely related, which will be discussed here, starting with the persistent dimerization.

Ti K-edge EXAFS shows that Ti forms dimers at all dopings in Li$_2$Ti$_x$Ru$_{1-x}$O$_3$, which is unexpected. This behavior contrasts that of Mn, which does not form dimers as a dopant in Ru site of Li$_2$RuO$_3$~\cite{Yun2021}. Iridium (Ir) was studied as another dopant for Li$_2$RuO$_3$, and was also found not to form dimers, when inserted in the Ru site~\cite{Lei2014}.
One possible explanation for the Ti-Ru bond in our samples may be the so-called early-late heterobimetallic bond, which has been previously observed in organometallics~\cite{Kabashima1999,Gade1999,Legendre2000,Legendre2003,Fujita2004}. It is a direct metal-metal bond formed between transition metal-ions of type $d^0$ (early) and $d^4/d^5$ (late)~\cite{Kabashima1999,Gade1999,Fujita2004}. The Ti-Ru bond distance in these cases is found to be close to 2.55 \textup{\AA}~\cite{Kabashima1999,Baranger1993}, close to the Ti-Ru dimer bond lengths observed in Fig.~\ref{fig:Ti}. This early-late heterobimetallic bond is also known to be orbital-selective~\cite{Lachguar2024,Cooper2012,Baranger1993,Baranger1994,Hanna1995,Hanna1996,Fulton2000}, much like the Ru-Ru molecular-orbit formation, which may explain the directionality of the dimers, that is, why only certain bonds dimerize. At high doping, the decrease in dimer presence results from the increased probability of Ti-Ti neighbours, which would not form such dimers and prefer $a_1$-type neighbours, as discussed in Sec.~\ref{sec:exafs}. 

\begin{figure}[htbp]
\centering
\includegraphics[width=0.5\textwidth]{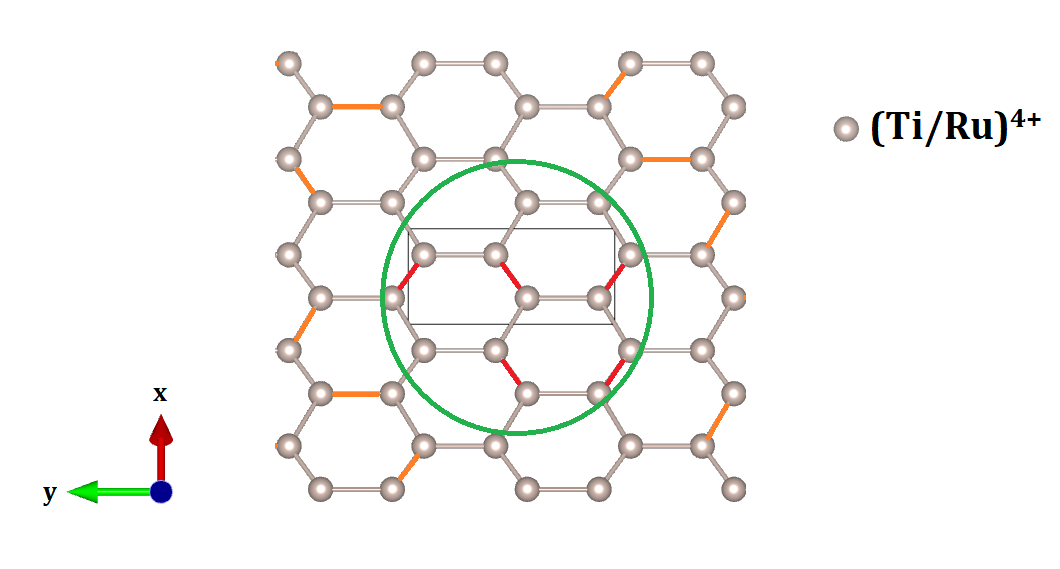}
\caption{\label{fig:cluster}Visualization of a proposed `dimer-cluster', with its boundary shown using the green circle. The red bonds indicate dimerized neighbours. The diameter of the green circle is the dimer-correlation length, an estimate of the coverage of the cluster. The dimer pattern of the in-plane hexagons is correlated within the cluster at all temperatures. The `inter-cluster' correlation breaks down at higher temperatures, with uncorrelated dimers shown in orange outside the cluster (green circle). The monoclinic unit cell is shown with the solid black line. Schematic representation made using the VESTA software~\cite{Momma2011}.}
\end{figure}

A previous study of Mn-doping of Li$_2$RuO$_3$~\cite{Yun2021} showed that the transition temperature monotonically decreased with an increase in Mn-doping, which is in sharp contrast with the non-monotonic behavior observed in the current study. It is also interesting to note that once Ti-doping starts to suppress $T_c$, it is almost twice as efficient as Mn. The (negative) slope of the Mn-doping dependence of $T_c$ is $\sim 550$~K, while that of Ti-doping is $\sim1050$~K (dashed line in Fig.~\ref{fig:XRD2}. 

In order to understand the non-monotonic $T_c$ behaviour and reconcile it with the neighbour distance trend, we model the system at high temperature as `dimer-clusters', whose size is given by the dimer-correlation length from the PDF results, Fig.~\ref{fig:PDF-corr}. The dimer-cluster size is roughly  $\approx 10$ \AA $ $ for the undoped sample, based on Fig.~\ref{fig:PDF-corr} and a previous study~\cite{Kimber2014}. These dimer clusters nucleate into long-range dimer order below $T_c$. When a small amount of Ru is replaced with Ti, no drastic change is expected since single Ti will stay as a dimer with one of the Ru neighbors, which may explain the insensitivity of $T_c$ to doping for $x < 0.2$. The dimer order will be affected when a substantial amount of Ti-Ti bonds replace Ru-Ru or Ru-Ti bonds, the Ti-Ti bond will separate dimer clusters and suppress nucleation of large distorted domains.


A heuristic picture of the dimer-cluster model is illustrated in Fig.~\ref{fig:cluster}. The dimer-cluster, inside the green circle, is made up of `correlated' dimer pattern with the red bonds indicating the dimerized neighbours (See Fig.~\ref{fig:s2}). Different dimer clusters, made up with the orange dimers are found outside the green circle. This averages out to the un-dimerized structure when considering all clusters outside the green boundary, leading to the high $R$ range fits in Fig.~\ref{fig:PDF-fit}. 

It is interesting to note that both $T_c$ and dimer correlation length seem to increase with dilute doping in Fig.~\ref{fig:XRD2} and Fig.~\ref{fig:PDF-corr}. Although the large error bars mean this result should be considered cautiously, when combined with the observation of Ti dimers in Fig.~\ref{fig:Ti}, this suggests that the enhancement of the cluster size may be due to Ti participating in dimerization. As Ti-Ru dimer bonds form, they can act as a `nucleation' point around which the dimer pattern `freezes'. These larger clusters might increase the energy cost of breaking `inter-cluster' correlation, leading to the slight enhancement of transition temperature ($T_c$) at low doping.

Our experimental observations raise an interesting question about whether the dimer-clusters are static or dynamic. That is, whether a particular region of the crystal structure is always a dimer cluster, or whether it shifts dynamically. For Ti-doped samples, it is reasonable to expect a static dimer-cluster, which is nucleated by Ti dopants. This is also supported by the fact that the cluster size is more or less temperature and doping-independent. However, proliferation of static clusters could suggest glassy behavior, which is incompatible with the observed clear phase transition in this system. Clearly, further studies, such as dynamic PDF measurements \cite{Acosta2023}, could be useful for addressing the nature of phase transition in Ti-doped and undoped Li$_2$RuO$_3$.

\section{Summary and Conclusions}
We present a comprehensive investigation of the structural phase transition in Ti-doped Li$_2$RuO$_3$, using three different types of structural probes: X-ray diffraction (XRD) for average structure, and EXAFS and PDF methods for local structure. Using XRD, we investigated the structural phase transition as a function of temperature over a wide doping range, and constructed the transition temperature versus doping concentration phase diagram. We found that replacing Ru with Ti has very little impact on the structural transition temperature in the limit of dilute Ti concentration. Only above a critical concentration of Ti ($x>0.2$), more or less linear suppression of the transition temperature is observed.

Local structural studies confirm the existence of the locally distorted dimer structure above the transition temperature, originally reported for the undoped system, for all doping ranges studied. Two notable observations are 1) Ti-Ru dimer formation confirmed by Ti K-edge EXAFS data, and 2) a dimer cluster of about $10-15$ \AA $ $ in all samples. Our PDF data indicates that such dimer clusters are remarkably robust and found in all samples in the high-temperature region, with little variation in their size.

The dimer-cluster model can qualitatively explain the observed phase diagram, since the dimer-clusters containing about 10 Ru-Ru bonds are unaffected by the low-level Ti-doping, and the transition temperature is also largely unaffected. Above the structural transition temperature, uncorrelated dimer clusters would average out to show an undistorted `global' crystal structure. 
The Ti-Ru dimer formation is attributed to the so-called early-late heterobimetallic bond, although further calculations would be desired to gain a quantitative understanding of the origin of Ti-Ru dimerization. In conclusion, our study demonstrates that the characterization of both `global' and local crystal structures can be a powerful tool that can be applied to the study of other materials with structural phase transitions. 

%

\end{document}